\documentclass[aps,preprint,showpacs,preprintnumbers,amsmath,amssymb]{revtex4}
\usepackage{amsmath,mathrsfs,amsbsy,color,graphicx,bm,amsthm,amsfonts}
\usepackage{units}
\usepackage{bbm}
\usepackage{times}
\usepackage{dcolumn}
\usepackage{mathrsfs}
\usepackage{amsmath,amssymb,epsfig}
\usepackage{amsmath}
\newcommand{\udots}{\mathinner{\mskip1mu\raise1pt\vbox{\kern7pt\hbox{.}}
\mskip2mu\raise4pt\hbox{.}\mskip2mu\raise7pt\hbox{.}\mskip1mu}}
\begin{document}
\title{Nonlocal coherence harvesting  from  quantum vacuum  }
\author{Rui-Di Wang,  Shu-Min Wu\footnote{smwu@lnnu.edu.cn (corresponding author)}, Xiao-Li Huang\footnote{ huangxiaoli1982@foxmail.com }}
\affiliation{Department of Physics, Liaoning Normal University, Dalian 116029, China
}


\begin{abstract}
It is well known that nonlocal coherence reflects nonclassical correlations better than quantum entanglement. Here, we analyze nonlocal coherence harvesting from the quantum vacuum to particle detectors adiabatically interacting with a quantum scalar field in Minkowski spacetime. We find that the harvesting-achievable separation range of  nonlocal coherence is  larger than that of quantum entanglement.
As the energy gap grows sufficiently large, the detectors harvest less quantum coherence, while the detectors could extract more quantum entanglement from the vacuum
state. Compared with the linear configuration and the scalene configuration, we should choose the model of equilateral triangle configuration to harvest tripartite coherence from the vacuum. Finally, we find a monogamous relationship, which means that tripartite $\mathrm{l_{1}}$-norm of coherence is essentially bipartite types.

\end{abstract}

\vspace*{0.5cm}
\maketitle

\section{Introduction}
Quantum coherence arises from the superposition of quantum states and is a basic characteristic of quantum physics \cite{L1}. Generating and maintaining quantum coherence is one of the necessary prerequisites for quantum information processing tasks \cite{L2}. Despite its importance, quantum coherence received increasing attention until Baumgratz $et$ $al.$ successfully applied the information-theoretic quantification of quantum coherence and introduced the measures based on the $\mathrm{l_{1}}$-norm and the relative entropy of all the potential metrics \cite{L3,L47}. Similar to quantum entanglement, quantum coherence is a significant quantum resource that plays a critical role in the fields of quantum biology, quantum metrology, quantum thermodynamic, and quantum computation \cite{L4,L5,L6,L7,L8,L9,L10,L11,L12,L13}. Particularly, the  phenomena of quantum entanglement can be attributed to the nonlocal superposition of quantum states.
It was confirmed that quantum coherence in a quantum system could be converted to quantum entanglement without providing further quantum coherence, meaning that quantum entanglement monotone can induce quantum coherence monotone of
quantum state \cite{L14}. Although the relationship between quantum coherence and entanglement has been further studied, it is still an open question worth investigating.

It has long been recognized that the vacuum states of a free quantum field theory in Minkowski space are highly entangled across spacelike regions \cite{L15}. Employing algebraic methods, Summers and Werner proved that correlations between field observables across spacelike regions can maximally violate Bell inequality \cite{L16,L17,L18,L19}. Subsequently, it was realized that this vacuum entanglement could be ``harvested" via particle detectors, which are linearly coupled to the quantum field \cite{L20,L21,L22}. Two-level particle detectors, called Unruh-DeWitt (UDW) detectors \cite{L23,L24,L25}, are like a particle in a ``box" interacting with the field when the ``box" opens. These detectors passing through the quantum field will be excited, showing that the vacuum state can be considered as a quantum resource for quantum entanglement and has been tested in a wide range of scenarios \cite{L26,L27,L28,L29,L30,L31,L32,L33,L34,L35}. In addition, entanglement harvesting can be used to detect the influences of nontrivial spacetime structure on vacuum entanglement, such as spacetime curvature, nontrivial spacetime topology, Unruh effect, Hawking effect, and cosmological effects \cite{L36,L37,L38,L39,L40,L41,L42,L42-3,L42-4,L42-5,L42-7,L42-8,L42-9,L42-10,L42-11,L42-12,L42-13,L42-15}. Because the Minkowski vacuum is an interesting nonclassical correlation and many of its properties remain unclear, it is worth further exploration.

The research so far has mainly focused on entanglement harvesting of UDW detectors. However, nonlocal coherence is a better proxy for nonclassical correlation than quantum entanglement. Therefore, in this paper, we study nonlocal coherence harvesting protocol with two UDW detectors adiabatically interacting with a quantum scalar field in $(3 + 1)$-dimensional Minkowski spacetime. Compared with entanglement harvesting, we will obtain the novel properties of nonlocal coherence and the differences between them. We also study nonlocal coherence of three detectors, considering three configurations of equilateral triangular, linear, and scalene triangular for three detectors, respectively. We will try to explore the relationship between bipartite coherence and tripartite coherence of three  detectors.

The structure of the paper is as follows. In Sec. II, we briefly introduce the UDW model. In Sec. III, we study nonlocal coherence of two UDW detectors. In Sec. IV, we extend the relevant research to three  detectors. The last section is devoted to a brief conclusion.
\section{The Unruh-DeWitt model}
Without losing generality, the UDW model of a detector $D$ with the ground state $|0\rangle_{D}$ and the excited state $|1\rangle_{D}$ is regarded as a two-level quantum system. We consider two detectors labeled $A$ and $B$ that interact locally with the massless quantum scalar field $\phi[x_{D}(\tau)]$. Then the interaction Hamiltonian between the detectors and the field can be expressed as
\begin{eqnarray}\label{S1}
H_{D}(\tau)=\lambda\chi(\tau)[\mathrm{e}^{\mathrm{i}\Omega_D\tau}\sigma^{+}+\mathrm{e}^{-\mathrm{i}\Omega_D\tau}\sigma^{-}]\phi[x_{D}(\tau)],\quad D\in\{A,B\}.
\end{eqnarray}
Here, $\Omega_{D}$ is the energy gap of the detector, the proper time $\tau$ parameterizes the classical spacetime trajectory of the detector, $\lambda$ is the coupling strength, $\sigma^{+}=|1_{D}\rangle\langle0_{D}|$ and $\sigma^{-}=|0_{D}\rangle\langle1_{D}|$ denote the ladder operators of the SU(2) algebra, $\chi(\tau)$ is the switching function, and $\phi[x_{D}(\tau)]$ is the pullback of the field operator on detector $D$'s trajectory.  The field operator can be defined as \cite{L42-16,L42-17}
\begin{eqnarray}\label{S1-1}
\phi(x)=\int \mathrm{d}^n k (u_\textbf{\emph{k}}(x)a_\textbf{\emph{k}}+u^{*}_\textbf{\emph{k}}(x)a^{\dag}_\textbf{\emph{k}}),
\end{eqnarray}
where $n=3$, $u_\textbf{\emph{k}}(x)=\frac{1}{\sqrt{2|\textbf{\emph{k}}|(2\pi)^3}}\mathrm{e}^{-\mathrm{i}|\textbf{\emph{k}}|t+\mathrm{i}\textbf{\emph{k}}\cdot \textbf{\emph{x}}}$, and $\{u_\textbf{\emph{k}}(x), u^{*}_\textbf{\emph{k}}(x)\}$ is a complete set of solutions of the field equation of motion and orthogonal to the Klein-Gordon inner product. Then, the field operator can be rewritten as
\begin{eqnarray}\label{S1-2}
\phi(x)=\int \frac{\mathrm{d} k^3}{(2\pi)^{3/2}} \frac{1}{\sqrt{2|\textbf{\emph{k}}|}}(\mathrm{e}^{-\mathrm{i}|\textbf{\emph{k}}|t+\mathrm{i}\textbf{\emph{k}}\cdot \textbf{\emph{x}}}a_\textbf{\emph{k}}+\mathrm{e}^{\mathrm{i}|\textbf{\emph{k}}|t-\mathrm{i}\textbf{\emph{k}}\cdot \textbf{\emph{x}}}a^{\dag}_\textbf{\emph{k}}).
\end{eqnarray}

For simplicity, we assume that the coupling constants are all identical, and $\lambda\ll1$ means weak coupling. In the process of interaction, the unitary operator $U$ is a description of the time evolution of the detectors and field and is generated by the interaction Hamiltonian in Eq.(\ref{S1}) as follows
\begin{eqnarray}\label{S2}
U:&=&\mathcal{T}\exp\left(-\mathrm{i}\int_{\mathbb{R}} \mathrm{d}\tau\ H_{D}(\tau)\right) \
=1+(-\mathrm{i}\lambda)\int_{\mathbb{R}} \mathrm{d}\tau\ H_{D}(\tau)\nonumber\\
&&+\frac{(-\mathrm{i}\lambda)^{2}}{2}\int_{\mathbb{R}} \mathrm{d}\tau \int_{\mathbb{R}} \mathrm{d}\tau' \mathcal{T}H_{D}(\tau)H_{D}(\tau')+\mathcal{O}(\lambda^3).
\end{eqnarray}

Initially, the two detectors are prepared (as $\tau\rightarrow-\infty$) in the ground state $|0_{A}0_{B}\rangle$, and the field is in a vacuum state $|0_{M}\rangle$. The initial density matrix is thus $\rho_{0}=|0_{A}0_{B}\rangle\langle0_{A}0_{B}|\otimes|0_{M}\rangle\langle0_{M}|$. With the interaction Hamiltonian, the final state of the system is found to be
\begin{eqnarray}\label{S3}
\rho_{AB}={\rm Tr}_{\phi}[ U(|0_{A}0_{B}\rangle\langle0_{A}0_{B}|\otimes|0_{M}\rangle\langle0_{M}|)U^{\dag}].
\end{eqnarray}
By performing the standard perturbation theory \cite{L35,L37}, the final state of the system can be rewritten as a density matrix in the basis $\{|0_{A}\rangle\langle0_{B}|, |0_{A}\rangle\langle1_{B}|, |1_{A}\rangle\langle0_{B}|, |1_{A}\rangle\langle1_{B}|\}$,
\begin{eqnarray}\label{S4}
\rho_{AB}=\left(\!\!\begin{array}{cccccccc}
\rho_{11} & 0 & 0 & \rho_{14}\\
0 & \rho_{22} & \rho_{23} & 0\\
0 & \rho^{*}_{23} & \rho_{33} & 0\\
\rho^{*}_{14} & 0 & 0 & \rho_{44}\\
\end{array}\!\!\right)
=\left(\!\!\begin{array}{cccccccc}
1-P_{A}-P_{B} & 0 & 0 & X\\
0 & P_{B} & C & 0\\
0 & C^{*} & P_{A} & 0\\
X^{*} & 0 & 0 & 0\\
\end{array}\!\!\right)
+\mathcal{O}(\lambda^4),
\end{eqnarray}
where
\begin{eqnarray}\label{S5}
P_{D}:=\lambda^{2}\int\int{\mathrm{d}\tau \mathrm{d}\tau'\chi(\tau)\chi(\tau')\mathrm{e}^{-\mathrm{i}\Omega(\tau-\tau')}W(x_{D}(t),x_{D}(t'))}\ \quad   D\in\{A,B\},
\end{eqnarray}
\begin{eqnarray}\label{S6}
C:=\lambda^{2}\int\int{\mathrm{d}\tau \mathrm{d}\tau'\chi(\tau)\chi(\tau')\mathrm{e}^{-\mathrm{i}\Omega(\tau-\tau')}W(x_{A}(t),x_{B}(t'))},
\end{eqnarray}
\begin{eqnarray}\label{S7}
X:&=&-\lambda^{2}\int\int{\mathrm{d}\tau \mathrm{d}\tau'\chi(\tau)\chi(\tau')\mathrm{e}^{-\mathrm{i}\Omega(\tau+\tau')}}\nonumber\\
&&[\theta(t'-t)W(x_{A}(t),x_{B}(t'))+\theta(t-t')W(x_{B}(t'),x_{A}(t))].
\end{eqnarray}
Here, $W(x,x')=\langle0_{M}|\phi(x)\phi(x')|0_{M}\rangle$ is the vacuum Wightman function of the field, $\theta(t)$ represents the Heaviside's step function, $t$ is a common time, $P_{D}$ is called the transition probability, and the quantities $C$ and $X$ characterize nonclassical correlations. Because of the weak coupling, $\lambda\ll1$, after the unitary operator $U$ with respect to the Unruh-DeWitt model is performed the Dyson series expansion, we only keep to the second order terms. The density matrix and some computation we get contain $\lambda^{4}$ that is very small, so we can ignore the minimal quantities for which the exponent of $\lambda$ exceeds four.

In the following, to make the parameter space less complex, we make all detectors identical, with the same energy gap $\Omega\sigma$ and switching function $\chi(\tau)$. Thus, the two detectors will have the same transition probabilities
\begin{eqnarray}\label{S8}
P_{A}=P_{B}=P.
\end{eqnarray}

Especially, $\chi(\tau)=\exp[-\tau^{2}/(2\sigma^{2})]$ is the Gaussian switching function, which controls the duration of interaction via parameter $\sigma$. Therefore, the matrix elements can be specifically expressed as \cite{L451,L45}
\begin{eqnarray}\label{S9}
P_{D}=\frac{\lambda^{2}}{4\pi}[\mathrm{e}^{-\Omega^{2}\sigma^{2}}-\sqrt{\pi}\Omega\sigma {\rm erfc}(\Omega\sigma)],
\end{eqnarray}
\begin{eqnarray}\label{S10}
C=\frac{\lambda^{2}}{4\sqrt{\pi}}\frac{\sigma}{L}\mathrm{e}^{-L^{2}/4\sigma^{2}}\bigg[{\rm Im} \left(\mathrm{e}^{\mathrm{i}\Omega{L}}{\rm erf} \left(\mathrm{i}\frac{L}{2\sigma}+\Omega\sigma\right)\right)-\sin(\Omega{L})\bigg]\in \mathbb{R},
\end{eqnarray}
\begin{eqnarray}\label{S11}
X=\frac{-\mathrm{i}\lambda^{2}}{4\sqrt{\pi}}\frac{\sigma}{L}\mathrm{e}^{-\sigma^{2}\Omega^{2}-L^{2}/4\sigma^{2}}{\rm erfc} \left(\mathrm{i} \frac{L}{2\sigma}\right),
\end{eqnarray}
where
\begin{eqnarray}\label{S12}
{\rm erf}(x):=\frac{2}{\sqrt{\pi}} \int^{x}_{0}\mathrm{d}t\mathrm{e}^{-t^{2}},\nonumber
\end{eqnarray}
\begin{eqnarray}\label{S13}
{\rm erfc}(x):=1-{\rm erf}(x).\nonumber
\end{eqnarray}

\section{Nonlocal coherence harvesting of two Unruh-DeWitt detectors}
A quantum state is considered coherent for a complete set of states if it can be represented as a nontrivial linear superposition of these states \cite{L2}.
The concept of quantum coherence comes directly from the superposition principle, which is one of the conceptual pillars of quantum theory \cite{L46}. In this paper, we examine nonlocal coherence harvesting for UDW detectors. Note that nonlocal coherence does not exist in a single subsystem. We briefly introduce two measures of quantum coherence: the $\mathrm{l_{1}}$-norm of coherence and the relative entropy of coherence (REC) \cite{L47}. In a reference basis $\{|i\rangle\}_{i=1,...,n}$ of a $n$-dimensional system, the $\mathrm{l_{1}}$-norm of quantum coherence can be defined as the sum of the absolute values of all the off-diagonal elements of the system density matrix $\rho$,
\begin{eqnarray}\label{S14}
C_{\mathrm{l_{1}}}(\rho)=\sum_{i\neq j}|{\rho}_{i,j}|,
\end{eqnarray}
and the measure of the REC can be written as
\begin{eqnarray}\label{S15}
C_{\rm{REC}}(\rho)=S(\rho_{\rm{diag}})-S(\rho),
\end{eqnarray}
where $S(\rho)$ indicates the von Neumann entropy of quantum state $\rho$, and $\rho_{\rm{diag}}$ is the state obtained from $\rho$ by removing all off-diagonal elements \cite{L47}.

Employing Eqs.(\ref{S4}) and (\ref{S14}), we can obtain the $\mathrm{l_{1}}$-norm of quantum coherence as
\begin{eqnarray}\label{S14-1}
C_{\mathrm{l_{1}}}(\rho_{AB})=2|C|+2|X|+\mathcal{O}(\lambda^4).
\end{eqnarray}
Next, we study the REC from the quantum vacuum. For this purpose, we need to calculate the eigenvalues of the density matrix $\rho_{AB}$. The density matrix of Eq.(\ref{S4}) has four nonzero eigenvalues
\begin{eqnarray}
&&\lambda_{1}=-C+P,\nonumber\\
&&\lambda_{2}=C+P,\nonumber\\
&&\lambda_{3}=\frac{1}{2}(1-2P-\sqrt{1-4P+4P^2+4|X|^2}),\nonumber\\
&&\lambda_{4}=\frac{1}{2}(1-2P+\sqrt{1-4P+4P^2+4|X|^2}).\nonumber
\end{eqnarray}
Thus, the REC of state $\rho_{AB}$ becomes
\begin{eqnarray}\label{S15-1}
C_{\rm{REC}}(\rho_{AB})=-\sum_{i}\alpha_{i}\log_{2}\alpha_{i}+\sum_{j=1}^4\lambda_{j}\log_{2}\lambda_{j}+\mathcal{O}(\lambda^4),
\end{eqnarray}
where $\alpha_{i}$ are the diagonal elements of $\rho_{AB}$ and $\lambda_{j}$ are nonzero eigenvalues of quantum state $\rho_{AB}$. Here, we emphasize that quantum coherence between two UDW detectors is nonlocal and does not exist in any subsystem.

In order to compare nonlocal coherence and quantum entanglement, we use the negativity to calculate quantum entanglement. Quantum entanglement between $A$ and $B$ in $\rho_{AB}$ is measured by the negativity $N(\rho_{AB})$, which can be defined as \cite{L48}
\begin{eqnarray}\label{S16}
N(\rho_{AB})=\frac{\parallel\rho_{AB}^{\mathrm{T}_{A}}\parallel-1}{2},
\end{eqnarray}
where $\mathrm{T}_{A}$ denotes the partial transpose of $\rho_{AB}$  concerning subsystem $A$. Here, $\parallel\cdot\parallel$ is the trace norm of a matrix, and $\parallel\rho_{AB}^{\mathrm{T}_{A}}\parallel-1$ equals two times of the sum of absolute values of negative eigenvalues of the matrix $\rho_{AB}^{\mathrm{T}_{A}}$. By using Eqs.(\ref{S4}) and (\ref{S16}), we can obtain the negativity $N(\rho_{AB})$ as
\begin{eqnarray}\label{S16-1}
N(\rho_{AB})=\max[0,|X|-P]+\mathcal{O}(\lambda^4).
\end{eqnarray}
So, the negativity $N(\rho_{AB})$ is a result of the competition
between  off-diagonal matrix element $X$ and  transition probabilities $P$, but it does not depend on the correlation term $C$.
From Eqs.(\ref{S14-1}) and (\ref{S15-1}), we can see that $\mathrm{l_{1}}$-norm of quantum coherence just depends on elements $C$ and $X$, while the REC depends on not only $C$ and $X$, but also $P$. Through the analysis of their analytic expressions, the nonlocal coherence can reflect the nonclassical correlation better than quantum entanglement.

\begin{figure}[htbp]
\centering
\includegraphics[height=1.8in,width=2.0in]{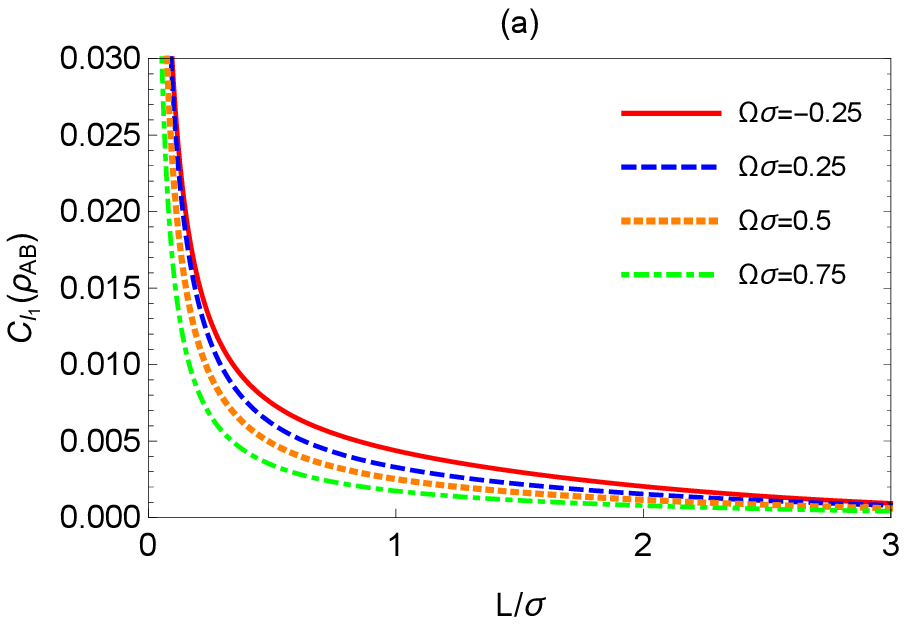}
\includegraphics[height=1.8in,width=2.0in]{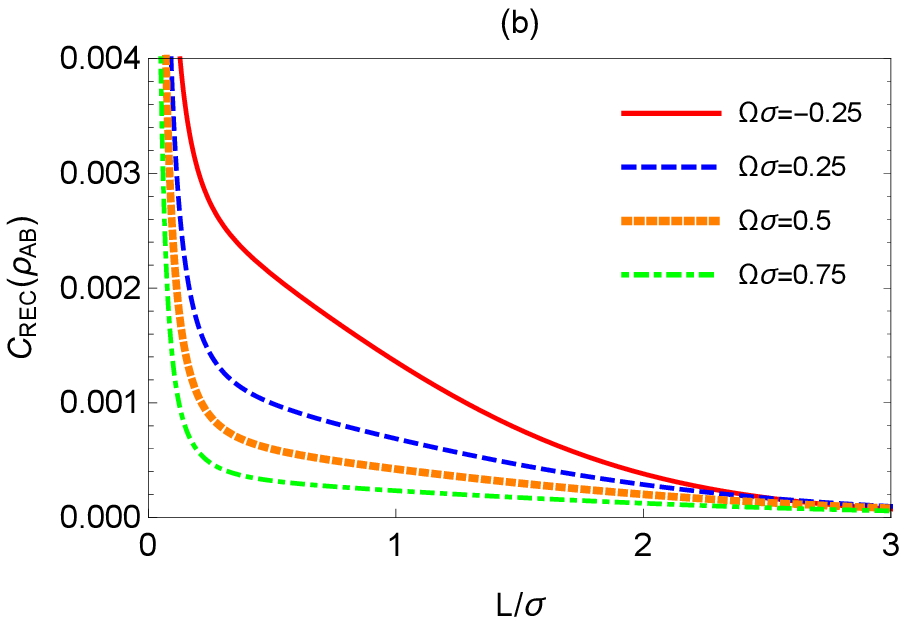}
\includegraphics[height=1.8in,width=2.0in]{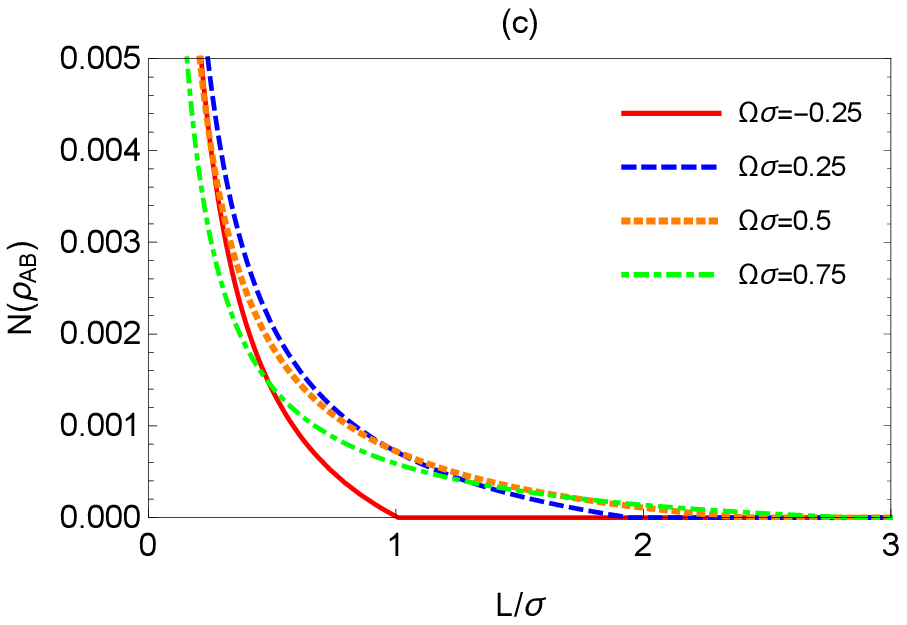}
\caption{Quantum coherence and entanglement of two UDW detectors as a function of the detector separation $L/\sigma$ for various values of the energy gap $\Omega\sigma$ with fixed $\lambda=0.1$. }\label{F1}
\end{figure}

In Fig.\ref{F1}, we plot quantum coherence (nonlocal coherence) and entanglement as a function of the detector separation $L/\sigma$ for different values of the energy gap $\Omega\sigma$. We find that quantum coherence first decreases and then tends to zero with the increase of the detector separation $L/\sigma$, meaning that the smaller detector separation is more conducive to nonlocal coherence harvesting of two UDW detectors. We also find that, for a fixed $L/\sigma$, the smaller energy gap $\Omega\sigma$ can harvest the larger quantum coherence, and nonlocal coherence harvesting for any energy gap $\Omega\sigma$ has a similar trend. However, from Fig.\ref{F1} (c) we can see that quantum entanglement first monotonically decreases and then suffers ``sudden death" with the increase of the detector separation $L/\sigma$. This means that the conditions for quantum entanglement harvesting are more demanding than nonlocal coherence harvesting. In other words, the harvesting-achievable separation range of  nonlocal coherence is  larger than that of quantum entanglement. We can also see that the variation of nonlocal coherence with the energy gap $\Omega\sigma$ is monotonous. Unlike nonlocal coherence, quantum entanglement is not monotonous with the energy gap $\Omega\sigma$, meaning that the smaller energy gap $\Omega\sigma$ is conducive to  nonlocal coherence  harvesting and may be detrimental to quantum entanglement  harvesting.
Note that the negative energy gap means that the atom transitions from the excited state to the ground state, as opposed to the positive energy gap, which is a transition from the initial ground state to the excited state \cite{L48-1,L48-2}.

\section{Nonlocal coherence harvesting of three Unruh-DeWitt detectors}
\begin{figure}[htbp]
\centering
\includegraphics[height=1.8in,width=2.0in]{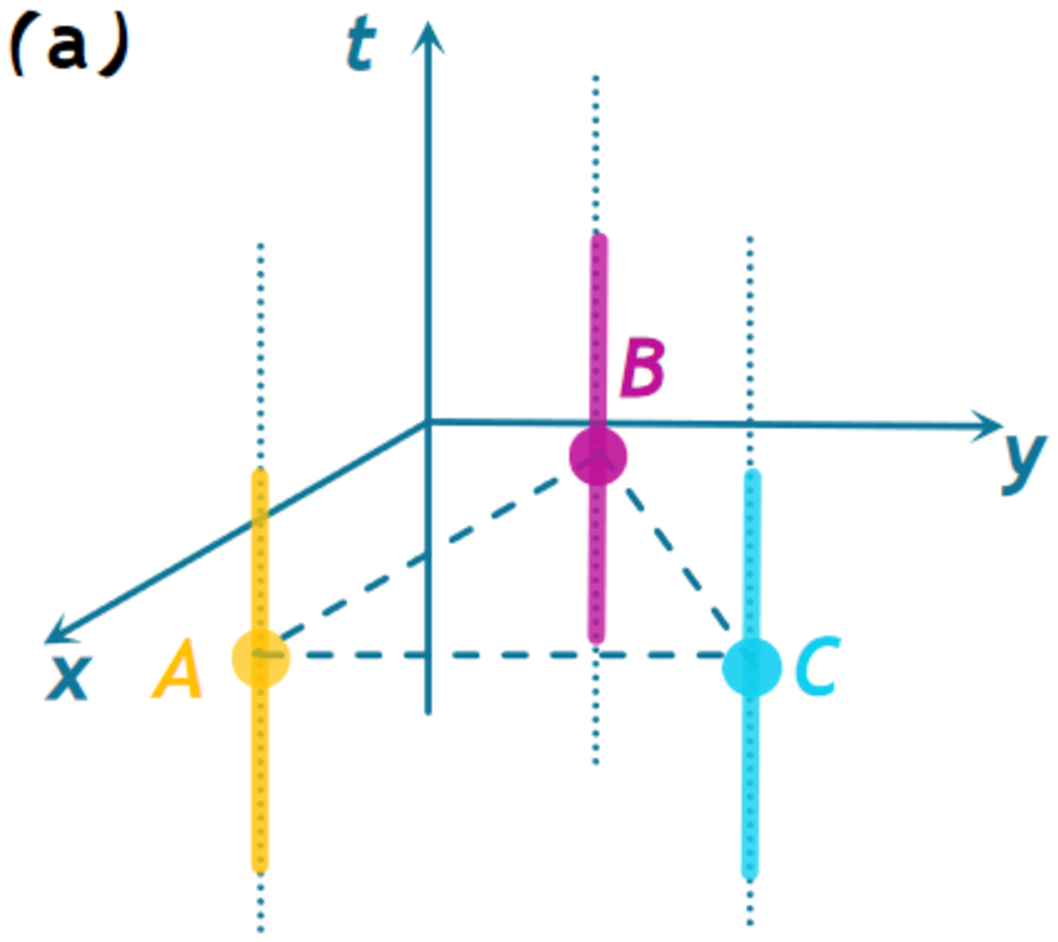}
\includegraphics[height=1.8in,width=2.0in]{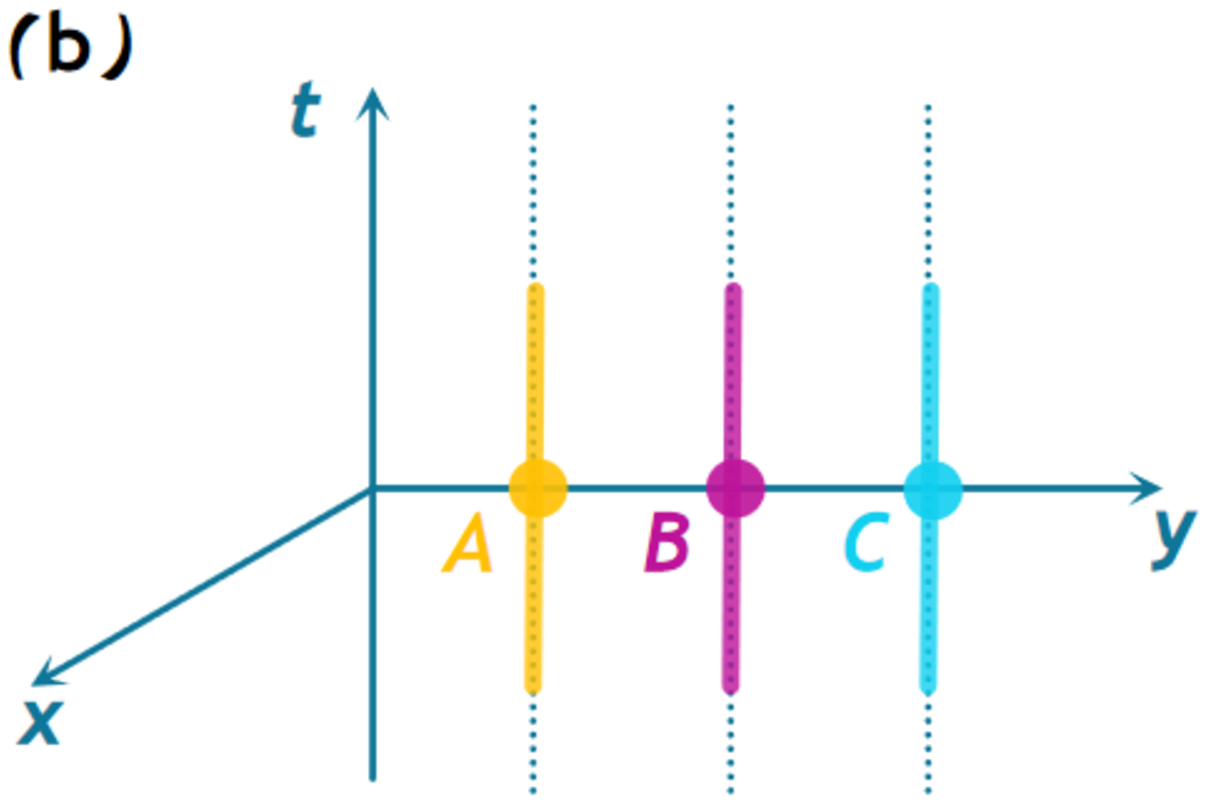}
\includegraphics[height=1.8in,width=2.0in]{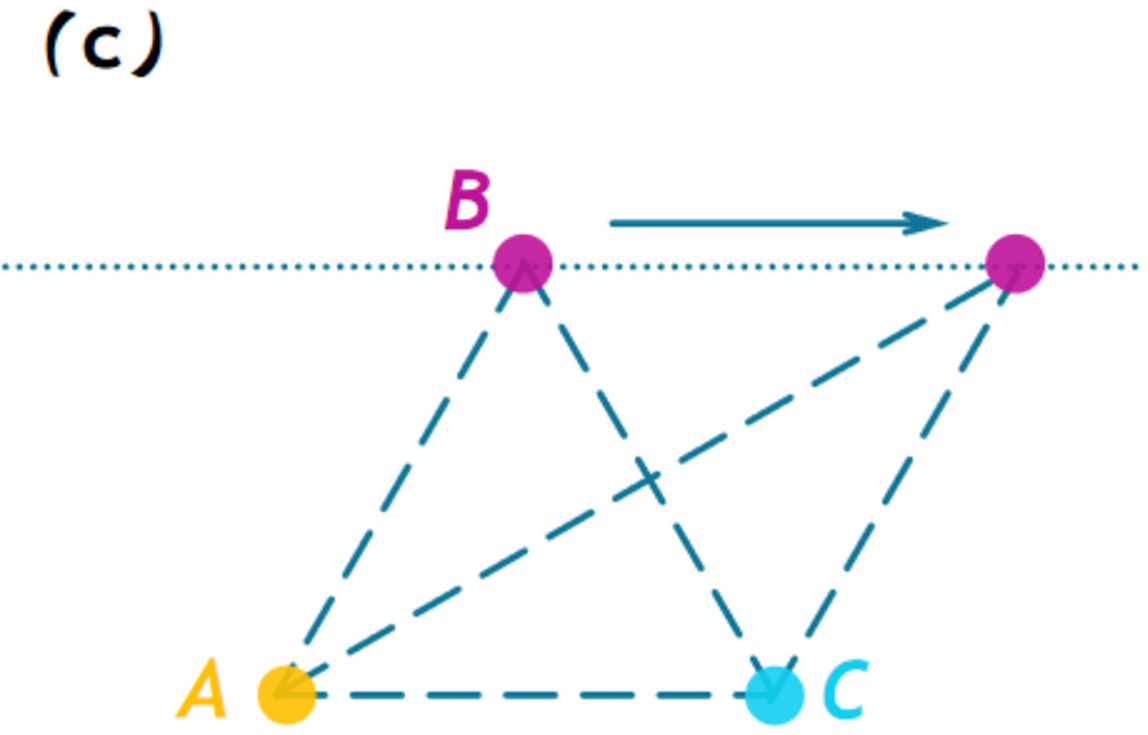}
\caption{Detector configurations we consider for the system of three detectors: (a) equilateral triangular, (b) linear, and (c) scalene triangular. (c) indicates the space disposition corresponding to (a) when $t=0$. \label{F2}}
\end{figure}

In the previous section, we  studied nonlocal coherence of two UDW detectors from the quantum vacuum. To investigate the relationship between tripartite coherence and bipartite coherence, we will consider the case of three UDW detectors \cite{L49}. Thus, using a similar method of two detectors, we can get the final state of the system of three detectors as
\begin{eqnarray}\label{S18}
\rho_{ABC} =\left(\!\!\begin{array}{cccccccc}
1-(P_{A}+P_{B}+P_{C}) & 0 & 0 & 0 & X_{BC} & X_{AC} & X_{AB} & 0\\
0 & P_{C} & C_{BC}  & C_{AC} & 0 & 0 & 0 & 0\\
0 & C_{BC}^* & P_B & C_{AB}  & 0  & 0 & 0 & 0\\
0 & C_{AC}^* & C_{AB}^*  & P_{A}  & 0 & 0 & 0 & 0\\
X_{BC}^* & 0 & 0 & 0 & 0 & 0 & 0 & 0\\
X_{AC}^* & 0 & 0 & 0 & 0 & 0 & 0 & 0\\
X_{AB}^* & 0 & 0 & 0 & 0 & 0 & 0 & 0\\
0 & 0 & 0 & 0 & 0 & 0 & 0 & 0\\
\end{array}\!\!\right)
+ \mathcal{O}(\lambda^4).\
\end{eqnarray}
Here, the basis vectors of the density matrix $\rho_{ABC}$ are $|0_{A}0_{B}0_{C}\rangle$, $|0_{A}0_{B}1_{C}\rangle$, $|0_{A}1_{B}0_{C}\rangle$, $|1_{A}0_{B}0_{C}\rangle$, $|0_{A}1_{B}1_{C}\rangle$, $|1_{A}0_{B}1_{C}\rangle$, $|1_{A}1_{B}0_{C}\rangle$, $|1_{A}1_{B}1_{C}\rangle$. We will take into account three types of detector configurations, which are shown in Fig.\ref{F2}. In the first configuration, we place the three detectors on each of the three vertices of an equilateral triangle and investigate how quantum entanglement varies with the detector separation $L/\sigma$. The second configuration is that the detectors are placed in a line of total length 2$L$, and the center detector is the same distance from the other two detectors, which changes the separation of the end detectors from the center one. Lastly, we consider a scalene triangular configuration. Detector $B$ is moved along the line parallel to the line connecting $A$ and $C$. The distance moved is denoted by $D$, where $A$ and $C$ remain at a fixed distance \cite{L26}.
\subsection{Equilateral triangle}
For the equilateral triangular arrangement, the system is viewed as symmetric. All distances between the identical detectors are the same. So we obtain the conditions as
\begin{eqnarray}\label{S19}
C_{BC}=C_{AC}=C_{AB} \equiv C,\quad X_{BC}=X_{AC}=X_{AB} \equiv X,\nonumber
\end{eqnarray}
and then the density matrix of Eq.(\ref{S18})  can be rewritten as
\begin{eqnarray}\label{S20}
\rho_{ABC} =\left(\!\!\begin{array}{cccccccc}
1-3P & 0 & 0 & 0 & X & X & X & 0 \\
0 & P & C & C & 0 & 0 & 0 & 0 \\
0 & C & P & C & 0 & 0 & 0 & 0\\
0 & C & C & P & 0 & 0 & 0 & 0 \\
X^* & 0 & 0 & 0 & 0 & 0 & 0 & 0 \\
X^* & 0 & 0 & 0 & 0 & 0 & 0 & 0 \\
X^* & 0 & 0 & 0 & 0 & 0 & 0 & 0 \\
0 & 0 & 0 & 0 & 0 & 0 & 0 & 0 \\
\end{array}\!\!\right)
+ \mathcal{O}(\lambda^4).\
\end{eqnarray}
Using Eqs.(\ref{S14}) and (\ref{S20}), we can get the $\mathrm{l_{1}}$-norm of quantum coherence of three detectors as
\begin{eqnarray}\label{S20-1}
C_{\mathrm{l_{1}}}(\rho_{ABC})=6|C|+6|X|+\mathcal{O}(\lambda^4),
\end{eqnarray}
and according to Eqs.(\ref{S15}) and (\ref{S20}), the REC becomes
\begin{eqnarray}\label{S20-2}
C_{\rm{REC}}(\rho_{ABC})=-\sum_{i}\alpha_{i}\log_{2}\alpha_{i}+\sum_{j=5}^{9}\lambda_{j}\log_{2}\lambda_{j}+\mathcal{O}(\lambda^4),
\end{eqnarray}
where $\alpha_{i}$ are the diagonal elements of $\rho_{ABC}$ of Eq.(\ref{S20}), and $\lambda_{j}$ are nonzero eigenvalues of quantum state $\rho_{ABC}$,
\begin{eqnarray}
&&\lambda_{5}=\lambda_{6}=-C+P,\nonumber\\
&&\lambda_{7}=2C+P,\nonumber\\
&&\lambda_{8}=\frac{1}{2}(1-3P-\sqrt{1-6P+9P^2+12|X|^2}),\nonumber\\
&&\lambda_{9}=\frac{1}{2}(1-3P-\sqrt{1-6P+9P^2+12|X|^2}).\nonumber
\end{eqnarray}
Note that quantum coherence of three UDW detectors is nonlocal and does not exist in any subsystem. Similar to the approach in the case of two detectors, we calculate the negativity of three detectors. The tripartite negativity of the three-qubit state $\rho_{ABC}$ is expressed as \cite{L50}
\begin{eqnarray}\label{S20-3}
&N(\rho_{ABC})&=(N_{A-BC}N_{B-AC}N_{C-AB})^{\frac{1}{3}}\nonumber\\
&&=\max[0,\frac{1}{2}\sqrt{C^2+|X|^2}-\frac{1}{2}C-P]+\mathcal{O}(\lambda^4).
\end{eqnarray}
\begin{figure}[htbp]
\centering
\includegraphics[height=1.8in,width=2.0in]{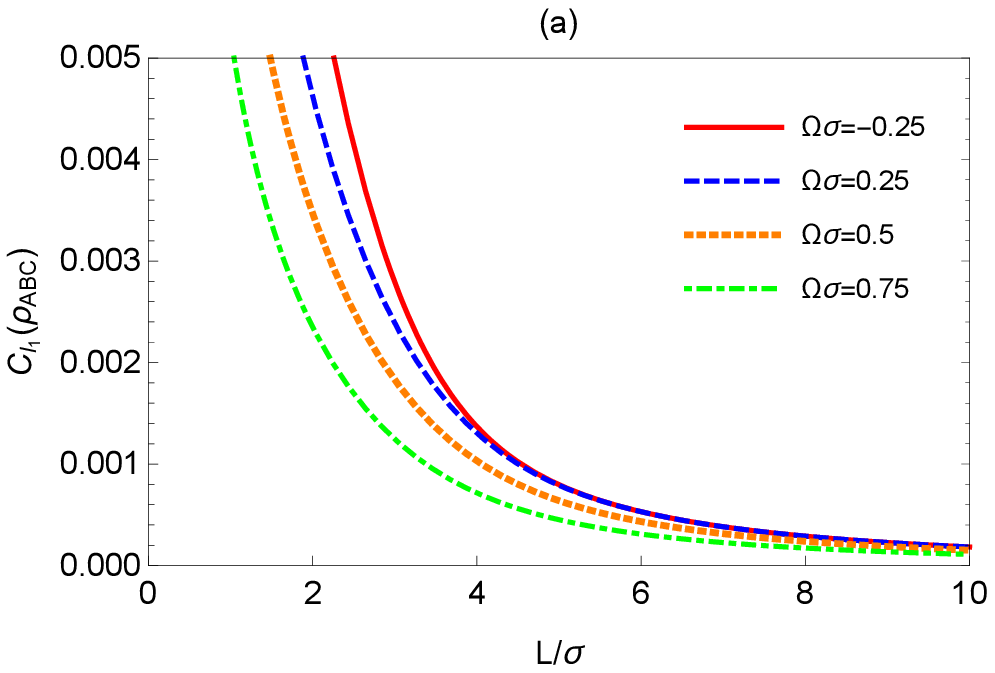}
\includegraphics[height=1.8in,width=2.0in]{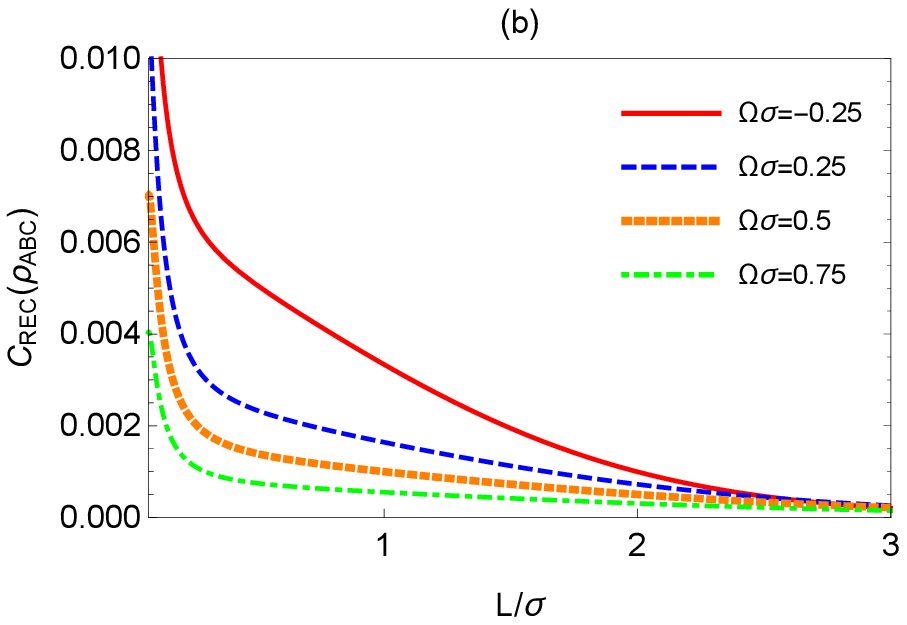}
\includegraphics[height=1.8in,width=2.0in]{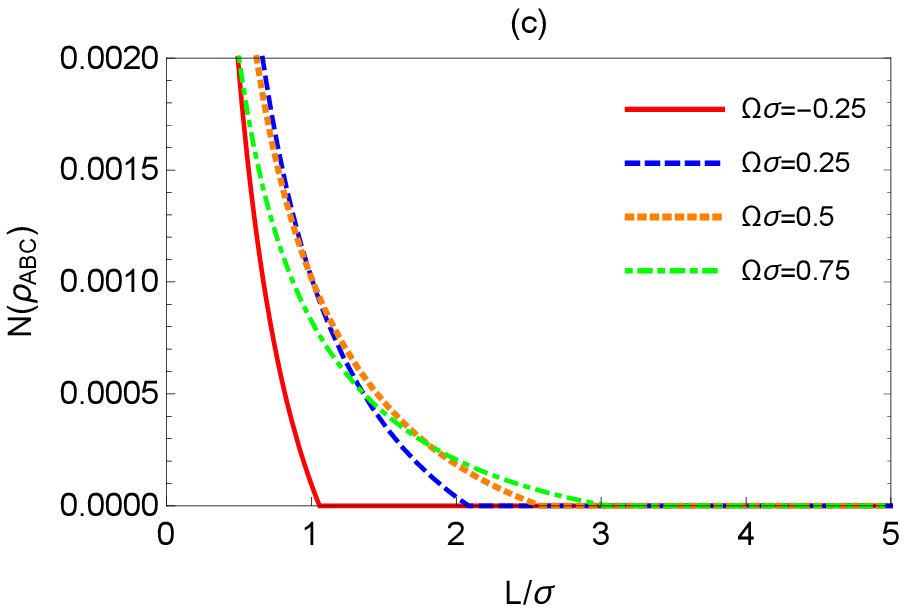}
\caption{Tripartite coherence and entanglement of three detectors in the equilateral triangular configuration as a function of the detector separation $L/\sigma$ for various values of the energy gap $\Omega\sigma$ with fixed $\lambda=0.1$. }\label{F3}
\end{figure}

In Fig.\ref{F3}, we plot tripartite coherence and entanglement as a function of the detector separation $L/\sigma$ for different $\Omega\sigma$ in the case of the equilateral triangular arrangement. From Fig.\ref{F3} (a) and (b), we see that tripartite coherence harvesting first monotonically reduces and then approaches zero with the growth of the $L/\sigma$. Tripartite coherence harvesting is larger for the smaller $\Omega\sigma$ with fixed  $L/\sigma$. From Fig.\ref{F3} (c), we also see that tripartite entanglement first monotonically reduces and then suffers ``sudden death" with increasing the $L/\sigma$. Tripartite coherence decreases with growing the $\Omega\sigma$, while tripartite  entanglement may increase with increasing the $\Omega\sigma$. Therefore, with the increase of the $L/\sigma$,  tripartite  entanglement may suffer ``sudden death" more prematurely for the smaller $\Omega\sigma$.

\subsection{Linear}
For the simple linear configuration, the system is seen as a straight line and detectors are placed at equal intervals. In this situation, $P_{D}$ will stay the same. But the matrix elements $X$ and $C$ will change. They are
\begin{eqnarray}\label{S21}
X_{BC}=X_{AB}\equiv X_{L},\quad C_{BC}=C_{AB}\equiv C_{L},\nonumber
\end{eqnarray}
and
\begin{eqnarray}\label{S21-1}
X_{AC}\equiv X_{2L},\quad C_{AC}\equiv C_{2L}.\nonumber
\end{eqnarray}
The density matrix of Eq. (\ref{S18}) has the form
\begin{eqnarray}\label{S22}
\rho_{ABC} =\left(\!\!\begin{array}{cccccccc}
1-3P & 0 & 0 &  0 & X_{L} & X_{2L} & X_{L} & 0 \\
0 & P & C_{L} & C_{2L} & 0 & 0 & 0 & 0 \\
0 & C_{L} & P & C_{L} & 0 & 0 & 0 &0\\
0 & C_{2L} & C_{L} & P & 0 & 0 & 0 & 0 \\
X_{L}^* & 0 & 0 & 0 & 0  & 0 & 0 & 0 \\
X_{2L}^* & 0 & 0 & 0 & 0 & 0 & 0 & 0 \\
X_{L}^* & 0 & 0 & 0 & 0 & 0 & 0 & 0 \\
0 & 0 & 0 & 0 & 0 & 0 & 0 & 0 \\
\end{array}\!\!\right)+\mathcal{O}(\lambda^4).
\end{eqnarray}
Employing Eqs.(\ref{S14}), (\ref{S15}) and (\ref{S22}), the $\mathrm{l_{1}}$-norm of quantum coherence and the REC are found to be
\begin{eqnarray}\label{S22-1}
C_{\mathrm{l_{1}}}(\rho_{ABC})=4|C_{L}|+2|C_{2L}|+4|X_{L}|+2|X_{2L}|+\mathcal{O}(\lambda^4),
\end{eqnarray}
\begin{eqnarray}\label{S22-2}
C_{\rm{REC}}(\rho_{ABC})=-\sum_{i}\alpha_{i}\log_{2}\alpha_{i}+\sum_{j=10}^{14}\lambda_{j}\log_{2}\lambda_{j}+\mathcal{O}(\lambda^4),
\end{eqnarray}
where $\alpha_{i}$ are the diagonal elements of $\rho_{ABC}$ of Eq.(\ref{S22}), $\lambda_{j}$ are nonzero eigenvalues of quantum state $\rho_{ABC}$,
\begin{eqnarray}
&&\lambda_{10}=P-C_{2L},\nonumber\\
&&\lambda_{11}=\frac{1}{2}(2P+C_{2L}-\sqrt{8C_{L}^2+C_{2L}^2}),\nonumber\\
&&\lambda_{12}=\frac{1}{2}(2P+C_{2L}+\sqrt{8C_{L}^2+C_{2L}^2}),\nonumber\\
&&\lambda_{13}=\frac{1}{2}(1-3P-\sqrt{1-6P+9P^2+8|X_{L}|^2+4|X_{2L}|^2}),\nonumber\\
&&\lambda_{14}=\frac{1}{2}(1-3P+\sqrt{1-6P+9P^2+8|X_{L}|^2+4|X_{2L}|^2}).\nonumber
\end{eqnarray}

\begin{figure}[htbp]
\begin{minipage}[t]{0.5\linewidth}
\centering
\includegraphics[width=3.0in,height=5.2cm]{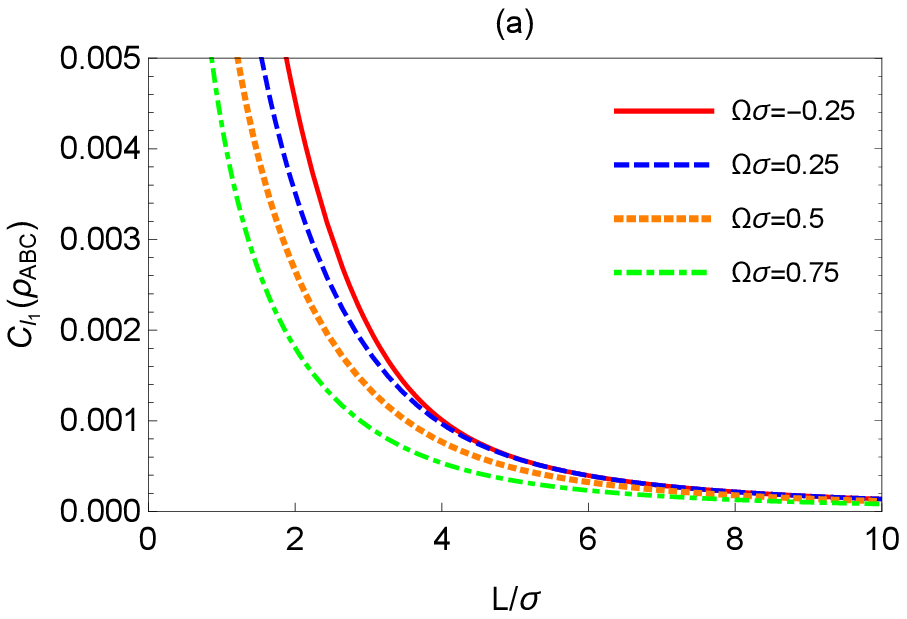}
\end{minipage}%
\begin{minipage}[t]{0.5\linewidth}
\centering
\includegraphics[width=3.0in,height=5.2cm]{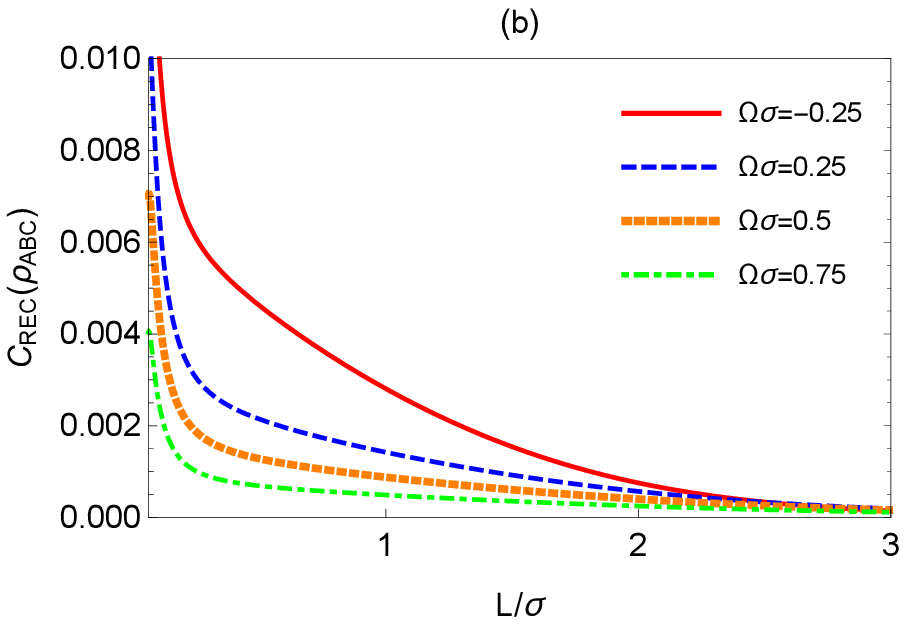}
\end{minipage}%
\caption{Quantum coherence of three detectors in the linear configuration as a function of the detector separation $L/\sigma$ for different energy gaps $\Omega\sigma$ with fixed $\lambda=0.1$.}\label{F4}
\end{figure}

Fig.\ref{F4} shows how the detector separation $L/\sigma$ influences  tripartite  coherence for different energy gaps $\Omega\sigma$ under the linear arrangement. Comparing Fig.\ref{F3} and Fig.\ref{F4}, we find that nonlocal coherence obtained in the equilateral triangle configuration is larger than in the linear configuration under the same conditions.

\subsection{Scalene triangle}
The scalene triangle can be regarded as a promotion of the equilateral triangle and the linear configuration. $P_{D}$ will remain identical. The matrix elements $X$ and $C$ will change when the distance between each pair of detectors is arbitrary. As a consequence, the density matrix of Eq.(\ref{S18}) can be expressed as
\begin{eqnarray}\label{S23}
\rho_{ABC} =\left(\!\!\begin{array}{cccccccc}
1-3P & 0 & 0 & 0 & X_{BC}  & X_{AC} & X_{AB} & 0 \\
0 & P & C_{BC}  & C_{AC} & 0 & 0 & 0 & 0 \\
0 & C_{BC}^* & P & C_{AB}  & 0  & 0 & 0 & 0 \\
0 & C_{AC}^* & C_{AB}^*  & P  & 0 & 0 & 0 & 0 \\
X_{BC}^* & 0 & 0 & 0 & 0 & 0 & 0 & 0 \\
X_{AC}^* & 0 & 0 & 0 & 0 & 0 & 0 & 0 \\
X_{AB}^* & 0 & 0 & 0 & 0 & 0 & 0 & 0 \\
0 & 0 & 0 & 0 & 0 & 0 & 0 & 0\\
\end{array}\!\!\right)+\mathcal{O}(\lambda^4).
\end{eqnarray}
Employing Eqs.(\ref{S14}) and (\ref{S23}), we obtain the $\mathrm{l_{1}}$-norm of quantum coherence as
\begin{eqnarray}\label{S23-1}
C_{\mathrm{l_{1}}}(\rho_{ABC})=2|X_{BC}|+2|X_{AC}|+2|X_{AB}|+2|C_{BC}|+2|C_{AC}|+2|C_{AB}|+\mathcal{O}(\lambda^4).
\end{eqnarray}
Since the analytic expression of the REC is complex, we cannot write it down specifically.
\begin{figure}[htbp]
\begin{minipage}[t]{0.5\linewidth}
\centering
\includegraphics[width=3.0in,height=5.2cm]{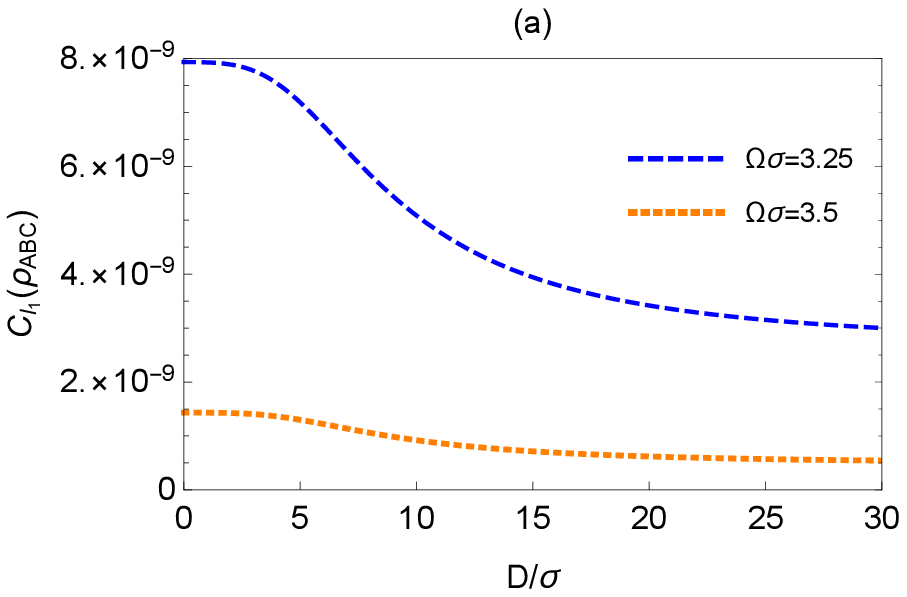}
\end{minipage}%
\begin{minipage}[t]{0.5\linewidth}
\centering
\includegraphics[width=3.0in,height=5.2cm]{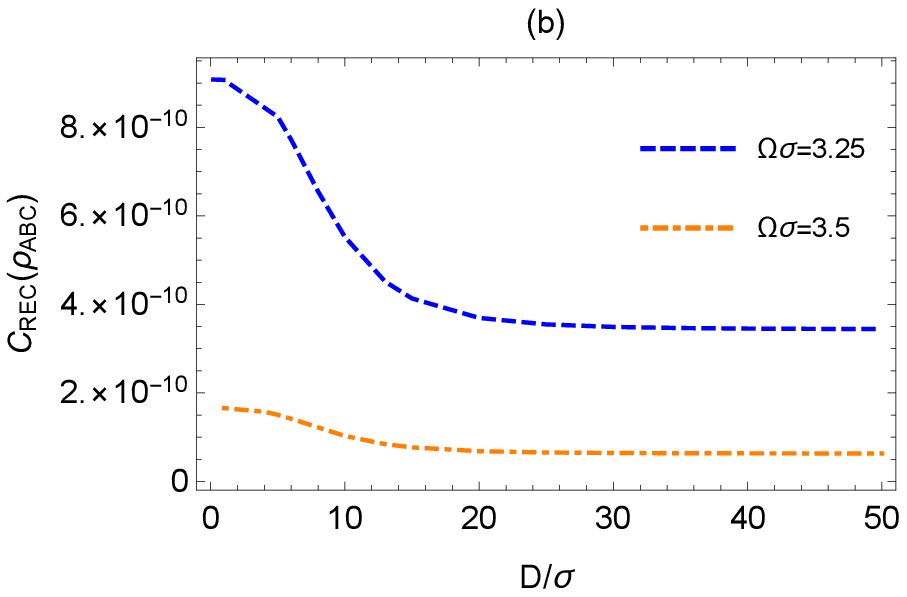}
\end{minipage}%
\caption{Quantum coherence of three detectors in the scalene configuration as a function of the detector separation $D/\sigma$ for different energy gaps $\Omega\sigma$ with $L_{AC}=7\sigma$. The coupling constant is set to $\lambda=0.1$.}\label{F5}
\end{figure}

In Fig.\ref{F5}, we plot nonlocal coherence of three UDW detectors of the scalene triangle arrangement as a function of the detector separation $D/\sigma$ for different energy gaps $\Omega\sigma$. It is shown that quantum coherence is a monotonically decreasing function of the separation $D/\sigma$. Combined with Fig.\ref{F3}-\ref{F5}, we can choose the model of equilateral triangle configuration to obtain tripartite coherence from the quantum vacuum.

Through calculation, for any type of detector configuration, we obtain the monogamy of the $\mathrm{l_{1}}$-norm of quantum coherence as
\begin{eqnarray}\label{S24}
C_{\mathrm{l_{1}}}(\rho_{AB})+C_{\mathrm{l_{1}}}(\rho_{BC})+C_{\mathrm{l_{1}}}(\rho_{AC})=C_{\mathrm{l_{1}}}(\rho_{ABC}).
\end{eqnarray}
From Eq.(\ref{S24}), we find that the $\mathrm{l_{1}}$-norm of coherence of three detectors is equal to the sum of the coherence of all two detectors. This indicates that the global tripartite $\mathrm{l_{1}}$-norm of coherence cannot be harvested from the quantum vacuum.

\section{Conclusion}
We have studied nonlocal coherence  harvesting of two or three Unruh-DeWitt detectors from the quantum vacuum. It is shown that bipartite coherence harvesting first decreases and then tends to zero as the detector separation increases.  However, bipartite entanglement harvesting first decreases and then suffers ``sudden death" with the growth of the detector separation. This suggests that  the harvesting-achievable separation range of nonlocal coherence is infinite, while the harvesting-achievable separation range of quantum entanglement is limited. We find that, with the decrease of the energy gap, bipartite coherence harvesting increases monotonically, while quantum entanglement harvesting is non-monotonically increasing.
Therefore, the results show that the smaller energy gap is beneficial for nonlocal coherence  harvesting but not necessarily  helpful  for quantum entanglement harvesting.

We have extended the investigation to  three types of detector configurations: equilateral triangular, linear, and scalene triangle.
Besides the similar properties observed in two Unruh-DeWitt detectors, some new characteristics have been found in three Unruh-DeWitt detectors.
Comparing nonlocal coherence of three configurations, we should select the model of equilateral triangle configuration for tripartite coherence harvesting from the quantum vacuum.
We obtain a monogamous relationship of tripartite $\mathrm{l_{1}}$-norm of coherence, which means that  tripartite coherence  equals the sum of all bipartite coherence.
In other words, there is no global  tripartite  $\mathrm{l_{1}}$-norm of coherence harvesting from the quantum vacuum.
Although both papers are devoted to the study of coherence and entanglement in the Unruh-DeWitt model, we will show the substantial differences between this paper and the reference \cite{L65}. On the one hand, the distance between the detectors in the reference \cite{L65} is very far, while the distance between the detectors in this paper is relatively close. On the other hand, the reference \cite{L65} discusses the effect of acceleration on coherence and entanglement when the two detectors are initially entangled, while this paper discusses the influence of detector separation on coherence and entanglement when two detectors are not entangled at the beginning.

\begin{acknowledgments}
This work is supported by the National Natural
Science Foundation of China (Grant Nos. 12205133 and 12075050), LJKQZ20222315 and JYTMS20231051.	
\end{acknowledgments}


\end{document}